\documentclass[11pt]{amsart}
\usepackage[margin=2cm]{geometry}                
\usepackage{graphicx}
\usepackage{amssymb}
\usepackage{epstopdf}
\title{Atmospheric characterisation of GJ1214b from transit $\&$ eclipse observations}
\author{$^1$P. Lavvas, $^2$S. Paraskevaidou, $^{1}$A. Arfaux}

\begin{document}
\maketitle
\begin{center}
{$^{1}$ Université Reims Champagne Ardenne, GSMA UMR CNRS 7331, Reims, France.\\
$^{2}$ Department of Physics, University of Crete, Heraklion, Greece.}
~~\\
~\\
\end{center}


\bf The atmospheric characterisation of GJ1214 b has so far remained uncertain due to the observed flatness of the transit spectra of this planet that is typically attributed to the presence of hazes or clouds in its atmosphere. Here we combine for the first time transit and eclipse observations obtained with JWST to benefit from both type of constraints and advance on the atmospheric characterisation of GJ1214 b. Our results reveal that photochemical hazes can be produced at high enough mass fluxes in the atmosphere of GJ1214 b to explain both type of observations. These hazes have a drastic impact on the atmospheric thermal structure, which has further ramifications on the emitted radiation of the planet, as well as, its Bond albedo. Clouds of KCL, NaCL and ZnS composition also form in this atmosphere but their opacity is too small to explain the observed flatness of the transit spectrum. We find that metallicities in the range 2000-3000x solar provide atmospheric structures that are closest to the observations for haze mass fluxes in the range of 1-3$\times$10$^{-11}$ g cm$^{-2}$ s$^{-1}$. Correspondingly the Bond albedo is within ~10-20$\%$. Moreover, sulfur photochemistry produces abundant OCS that has a detectable signature in the transit spectra and should be seaked for in future observations. Sulfur should also participate to the haze formation in this atmosphere, therefore optical properties of such compounds are needed.  \rm


GJ1214 b is a prototype sub-Neptune exoplanet. Characterising this type of environments is indispensable for understanding the long term evolution of planetary atmospheres and how atmospheric escape, secondary outgassing and other physical processes affect the observed distribution of planetary bodies \cite{Owen17, Kite20, Koskinen22}.  
The featureless transit spectrum, observed with HST/WFC3 in the near IR \cite{Kreidberg14}, limits any conclusions about the atmospheric composition, while other pre-JWST ground based \cite{Narita13, Nascimbeni15} and space borne \cite{Fraine13} observations across the spectrum, further support a globally flat spectrum and do not allow for any significant advancement on the atmospheric constraints. Therefore, the presence of heterogeneous components combined with a high metallicity atmospheric composition are typically called for explaining the observed spectral flatness \cite{Morley13, Morley15}. Such studies consider either condensate clouds \cite{Charnay15a, Gao18, Ohno18} or photochemical hazes \cite{Kawashima18, Adams19,Lavvas19} and evaluate the conditions required to explain the observations for each type of heterogeneous opacity. 

The thermal phase curve of GJ1214 b obtained with JWST/MIRI observations between 5 and 12 $\mu$m \cite{Kempton23} suggests a high metallicity environment covered by a thick and reflective layer of clouds or hazes that cause a large Bond albedo ($\sim$50$\%$). Moreover, analysis of the transit component of the JWST observations in tandem with previous transit spectra indicates that a strong production of photochemical hazes must take place in this atmosphere \cite{Gao23}. 
Here, we combine for the first time transit and eclipse observations to benefit from both type of constraints and evaluate if we can advance further the atmospheric characterisation of GJ1214 b. We find that the precision of the JWST observations provide stringent constraints on the atmospheric metallicity and the properties of the scattering particles.

We use a self-consistent 1D model to study the atmospheric properties of GJ1214 b. The model couples between the atmospheric chemistry in disequilibrium, the haze/cloud microphysics evaluated in bin mode, and the atmospheric thermal structure evaluated by radiative and convective energy transfer \cite{Arfaux22}. The advantage of such a description is that it provides a complete picture of the interplay among these three components and allows for a quick evaluation of the atmospheric conditions that can be directly compared with observations. We explore how different metallicity cases may affect the resulting spectrum of the planet in both transmission and emission. We study cases corresponding to metallicities between 100x and 10 000x solar \cite{Lodders19b} and for each metallicity case we consider scenarios  corresponding to clear (only gases) or hazy/cloudy atmospheres. We explore the microphysics of both hazes and clouds in tandem (see model description). Although the observations point away from a clear atmosphere we evaluate such a scenario for each metallicity case to understand how the thermal structure of the atmosphere would vary with respect to temperatures obtained for the hazy/cloudy atmosphere scenarios. 

\begin{figure}
\centering
\includegraphics[scale=0.45]{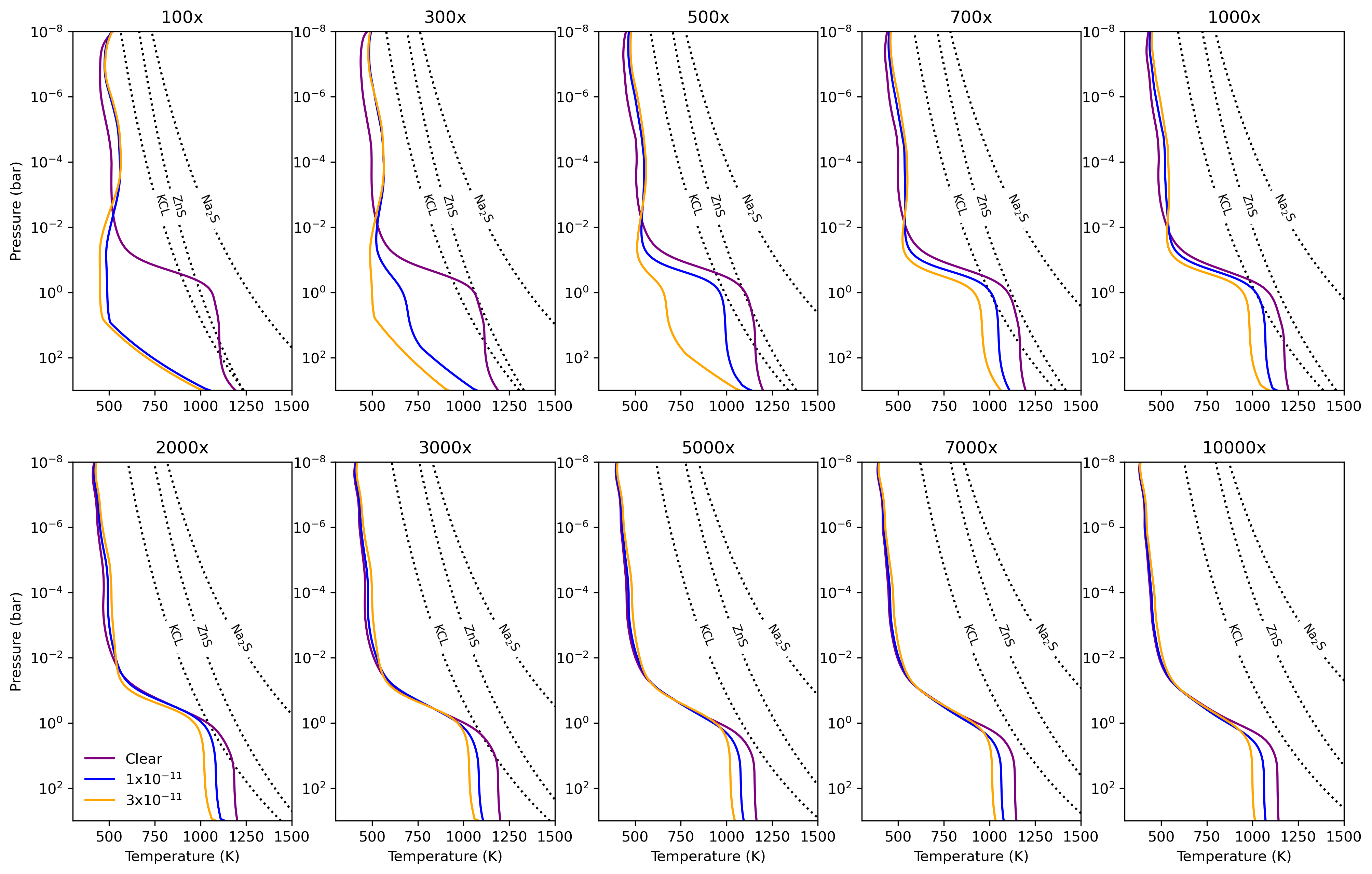}
\caption{Atmospheric temperature profiles for the different metallicity cases considered. Purple lines present the clear atmospheres, while blue and orange lines correspond to the 10$^{-11}$ and 3$\times$10$^{-11}$ g cm$^{-2}$s$^{-1}$ mass flux cases, respectively. Dotted lines present condensation curves \cite{Morley15}. }\label{temperature}
\end{figure}

Our simulated temperature profiles for clear atmospheres, qualitatively reveal a quasi-isothermal temperature near 1200 K in the deep atmosphere (1000 - 10 bar) followed by a drop in temperature at pressures between 10 bar and 10 mbar, before reaching a quasi-isothermal profile near 500 K at pressures up to 1 $\mu$bar (Fig.~\ref{temperature}). There are small quantitative changes with metallicity  from this general picture, since with increasing metallicity the atmospheric opacity rises. 
We additionally tested haze mass flux scenarios of 1$\times$10$^{-11}$ and 3$\times$10$^{-11}$ g cm$^{-2}$s$^{-1}$, also allowing for cloud formation by KCL, NaCL and ZnS condensation when the atmospheric temperature conditions allow it. The haze inclusion leads to heating of the upper atmosphere (p$<$100 mbar) and cooling of the atmosphere below, relative to the clear atmosphere temperature profiles. The magnitude of this effect decreases with increasing metallicity, as the relative contribution of gaseous opacity over that of the haze opacity increases. Nevertheless, the impact of hazes is observable for all simulated metallicity cases. 

\begin{figure}
\centering
\includegraphics[scale=0.5]{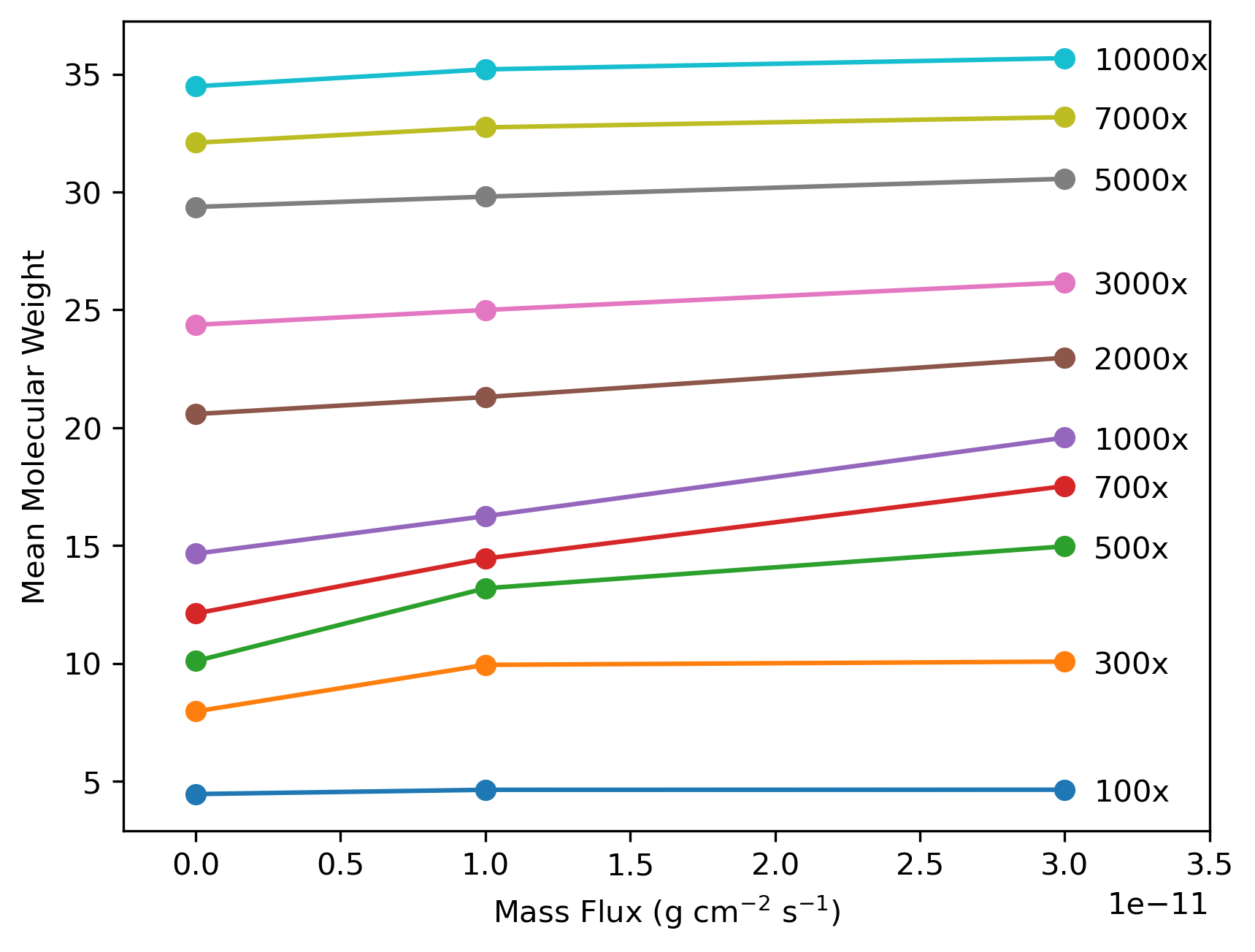}
\includegraphics[scale=0.5]{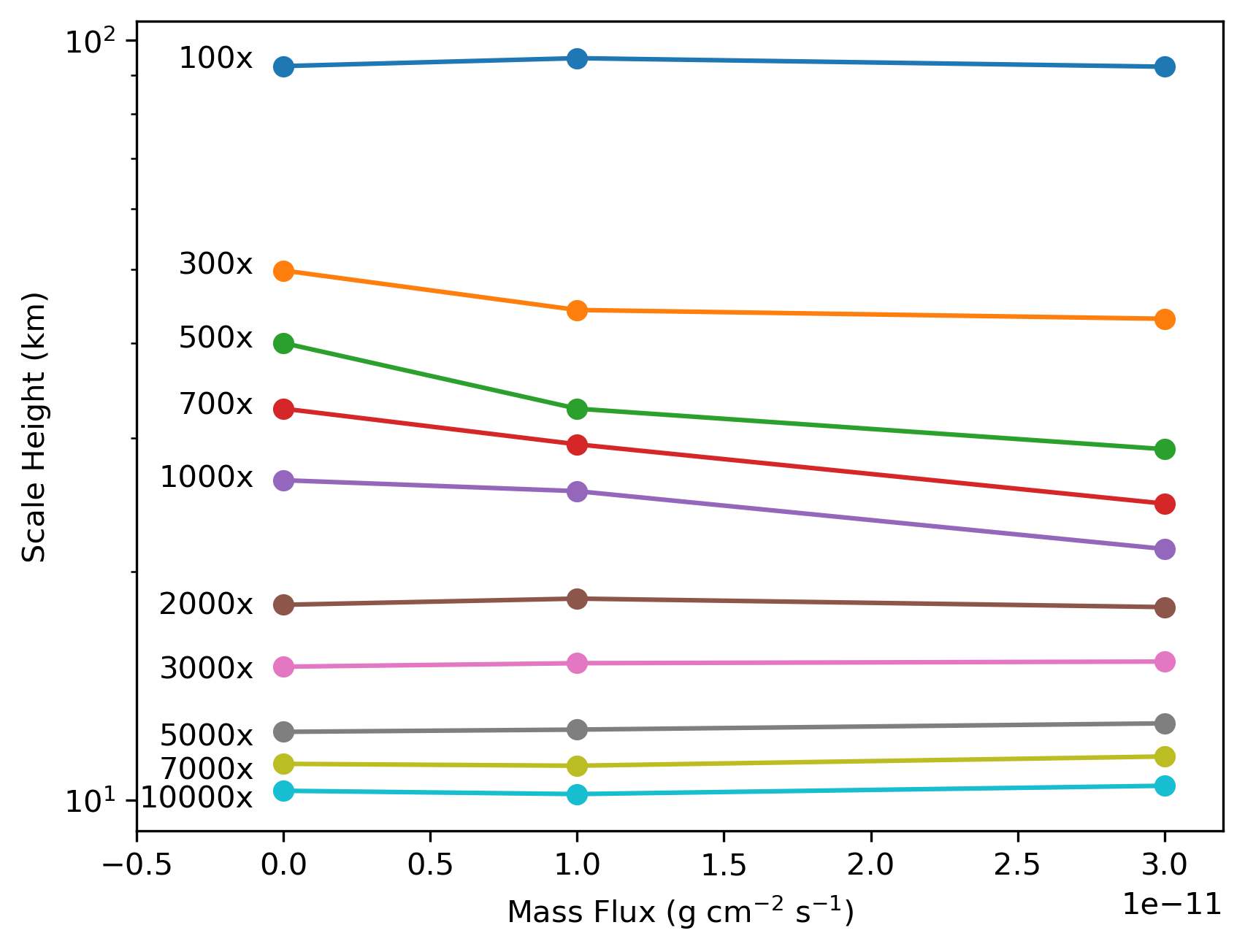}
\caption{Atmospheric mean molecular weight and scale height at the 1 mbar level evaluated under different metallicity and haze mass flux scenarios.}\label{mmw}
\end{figure}

The atmospheric cooling due to the haze propagates to the deep atmosphere, therefore affects the resulting chemical composition with implications for the atmospheric mean molecular weight and the resulting scale height (Fig.~\ref{mmw}). Haze inclusion mostly modifies the mean molecular weight at metallicities between 300x and 3000x solar as at these cases the temperature difference in the deep atmosphere is large enough to affect the distribution of the most abundant gas species. At lower metallicities, although temperature differences are significant the abundances of the gaseous species affected are not high enough to affect the dominance of the H$_2$/He envelope on the mean molecular weight. At higher metallicities, the impact of hazes on the temperature decreases, therefore, modifications in the chemical composition are smaller. 

\begin{figure}
\centering
\includegraphics[scale=0.45]{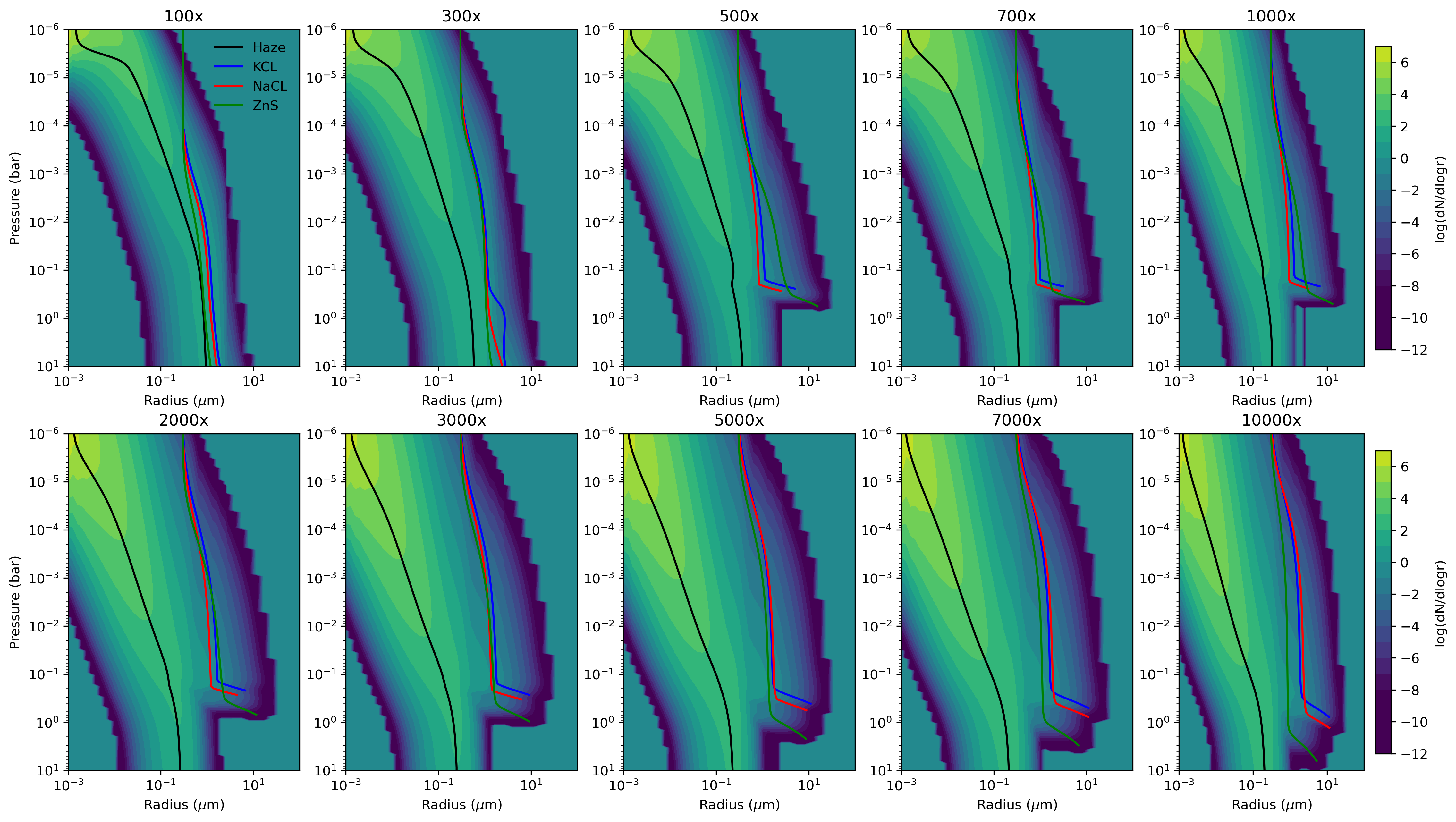}
\caption{Particle size distribution for hazes and clouds at different metallicity cases for the 10$^{-11}$ haze mass flux case. Contours present the cumulative particle number density and lines the average radius for the haze, KCL, NaCL and ZnS cloud particles.}\label{SD}
\end{figure}

The simulated haze particle distribution presents the typical growth from the formation region at 1 $\mu$bar with 1 nm radius particles to larger radii and lower densities at lower altitudes, due to coagulation (Fig.~\ref{SD}). Atmospheric metallicity affects the coagulation (through atmospheric temperature) and sedimentation velocity (through atmospheric viscosity) of the particles. The combined effect results in smaller average particle radius with increasing metallicity. For example, in the 1$\times$10$^{-11}$ g cm$^{-2}$s$^{-1}$ haze mass flux case the average haze particle size at 0.1 bar decreases from $\sim$0.7 $\mu$m at the 100x solar metallicity case to $\sim$0.07 $\mu$m at the 10 000x solar metallicity case, while the corresponding haze particle density increases from $\sim$10 cm$^{-3}$ to 600 cm$^{-3}$, between these metallicity limits.

\begin{figure}
\centering
\includegraphics[scale=0.45]{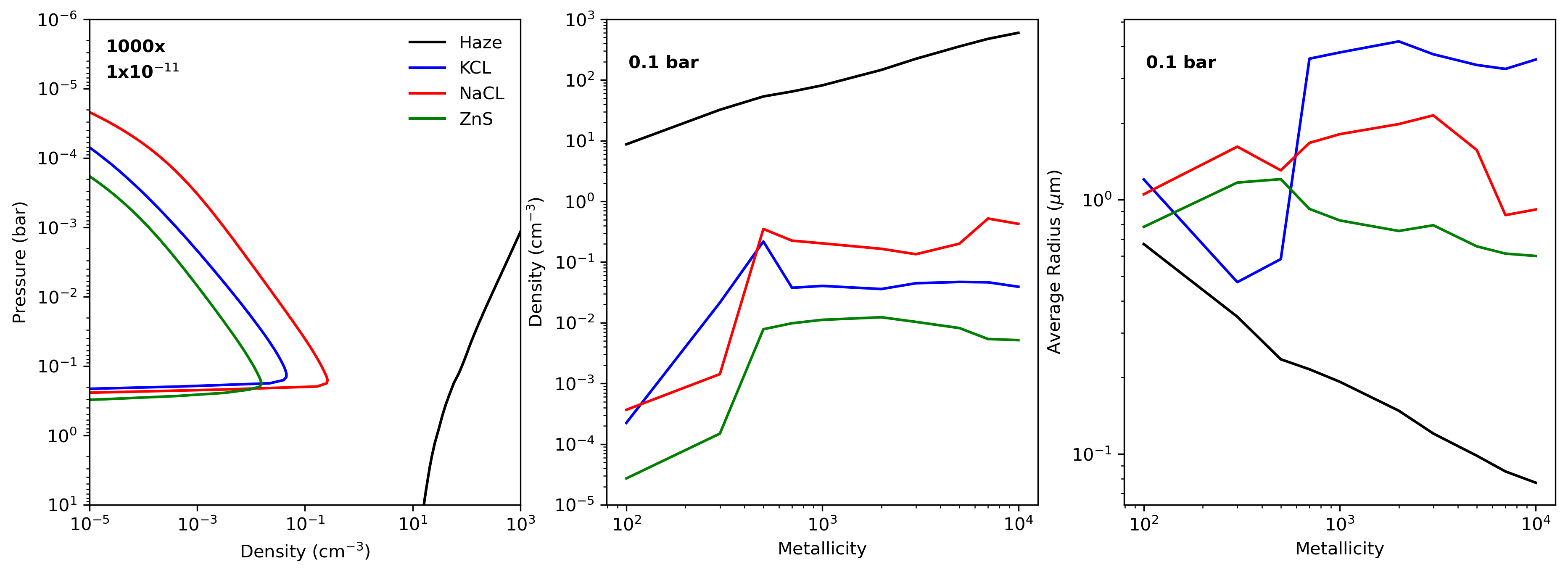}
\includegraphics[scale=0.45]{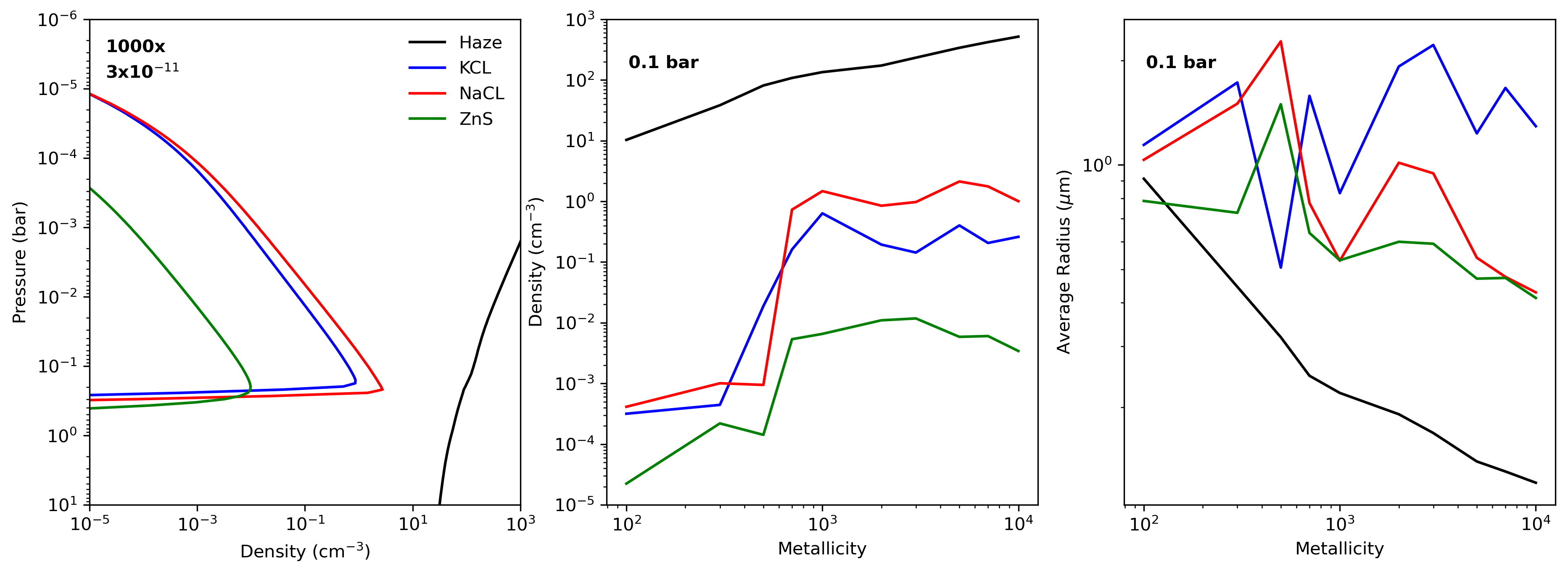}
\caption{Haze and cloud particle density profiles for the 1000x solar metallicity case (left) and average particle radius (middle) and density (right) at 0.1 bar, at different metallicity cases. The two rows corresponds to the 10$^{-11}$ (top) and the 3$\times$10$^{-11}$ g cm$^{-2}$s$^{-1}$ (bottom) haze mass flux cases. }\label{average_micro}
\end{figure}

Cloud formation depends strongly on the atmospheric temperature profile. For the 100x and 300x solar metallicity cases, the strong temperature drop in the deep atmosphere due to the haze limits cloud formation to very deep levels. However at higher metallicities, strong nucleation rates occur near 0.1 bar and cause the extension of the particle size distribution to radii larger than those obtained from haze coagulation only (Fig.~\ref{SD}). For the 10$^{-11}$ g cm$^{-2}$s$^{-1}$ haze mass flux case, average cloud radii at 0.1 bar are in the range of $\sim$0.5-5 $\mu$m and cloud particle densities have maxima at the order of 1 cm$^{-3}$ (Fig.~\ref{average_micro}). Among the cloud contributions, KCL and NaCL have similar characteristics in terms of particle size/density, and pressure range of formation, while the ZnS cloud formation peaks at slightly higher pressures with systematically lower cloud particle density compared to the other cloud types. For the higher haze mass flux case, cloud particle densities in principle should increase since there are more sites to nucleate, while the corresponding average cloud particle radii should decrease since the condensing mass is distributed to a larger number of particles. However, the resulting cloud properties demonstrate a variable behaviour with metallicity due to the implications of the haze on the thermal structure and further on the gas phase atmospheric composition that controls the abundance of the condensable species.

\begin{figure}
\includegraphics[scale=0.5]{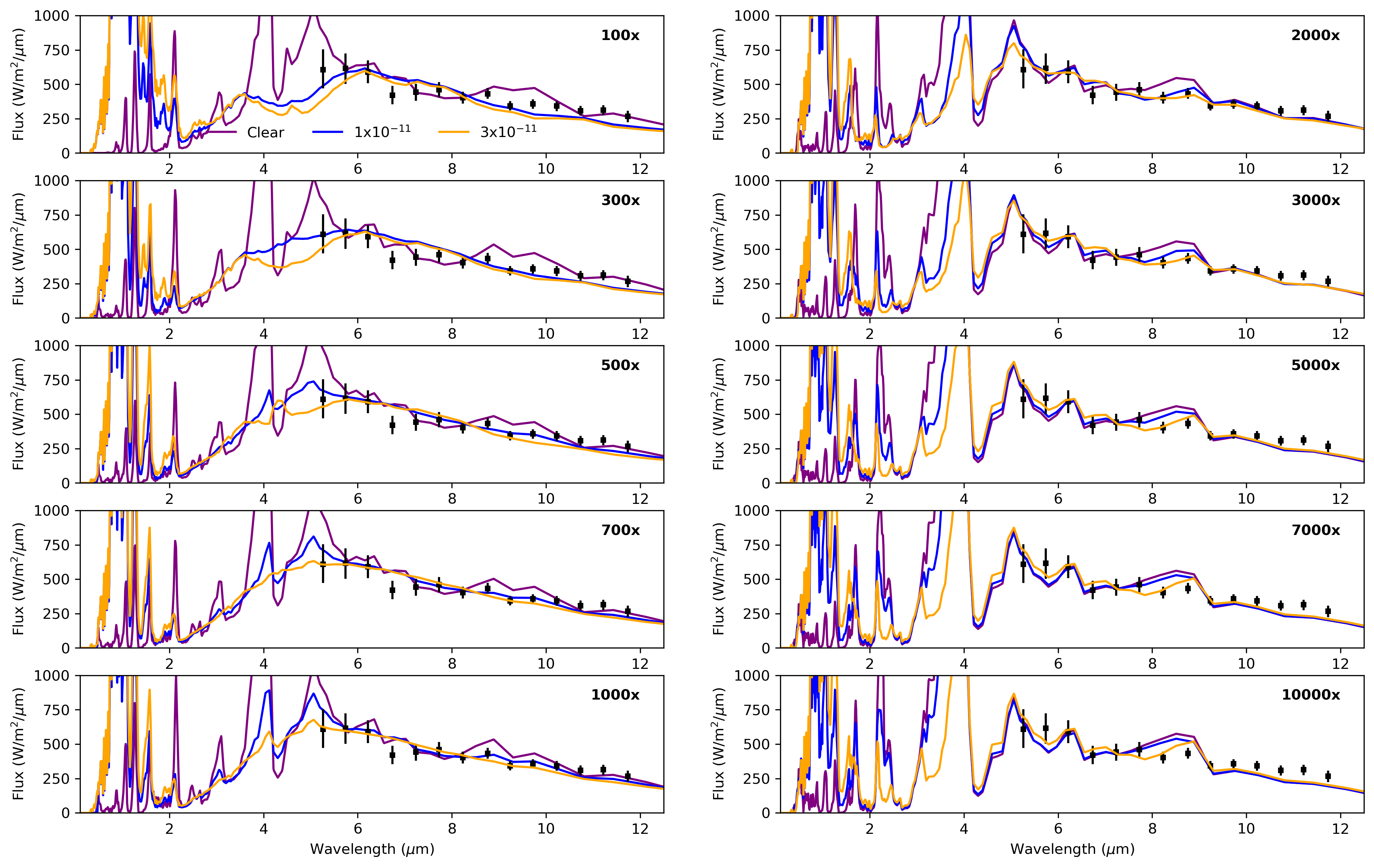}
\caption{Observed (points with error bars corresponding to one standard deviation) and simulated (lines) eclipse spectra of GJ 1214b for different metallicity scenarios (panels). Lines corresponds to clear atmosphere (purple), and the 10$^{-11}$ (blue) and 3$\times$10$^{-11}$ g cm$^{-2}$s$^{-1}$ (orange) haze mass flux scenarios. The simulated spectra include both stellar reflected and thermally emitted radiation.}\label{eclipse}
\end{figure}

\begin{figure}
\includegraphics[scale=0.5]{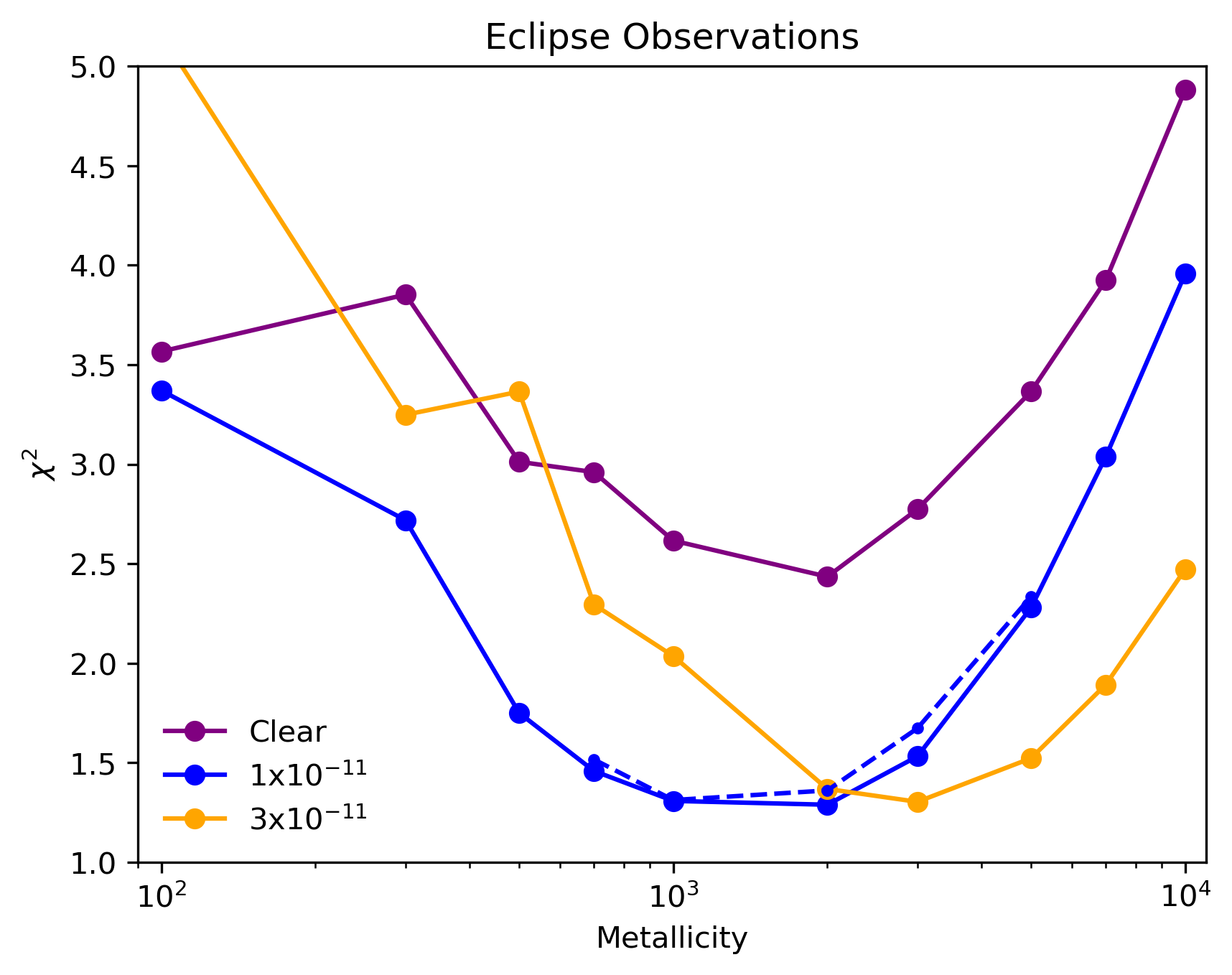}
\includegraphics[scale=0.5]{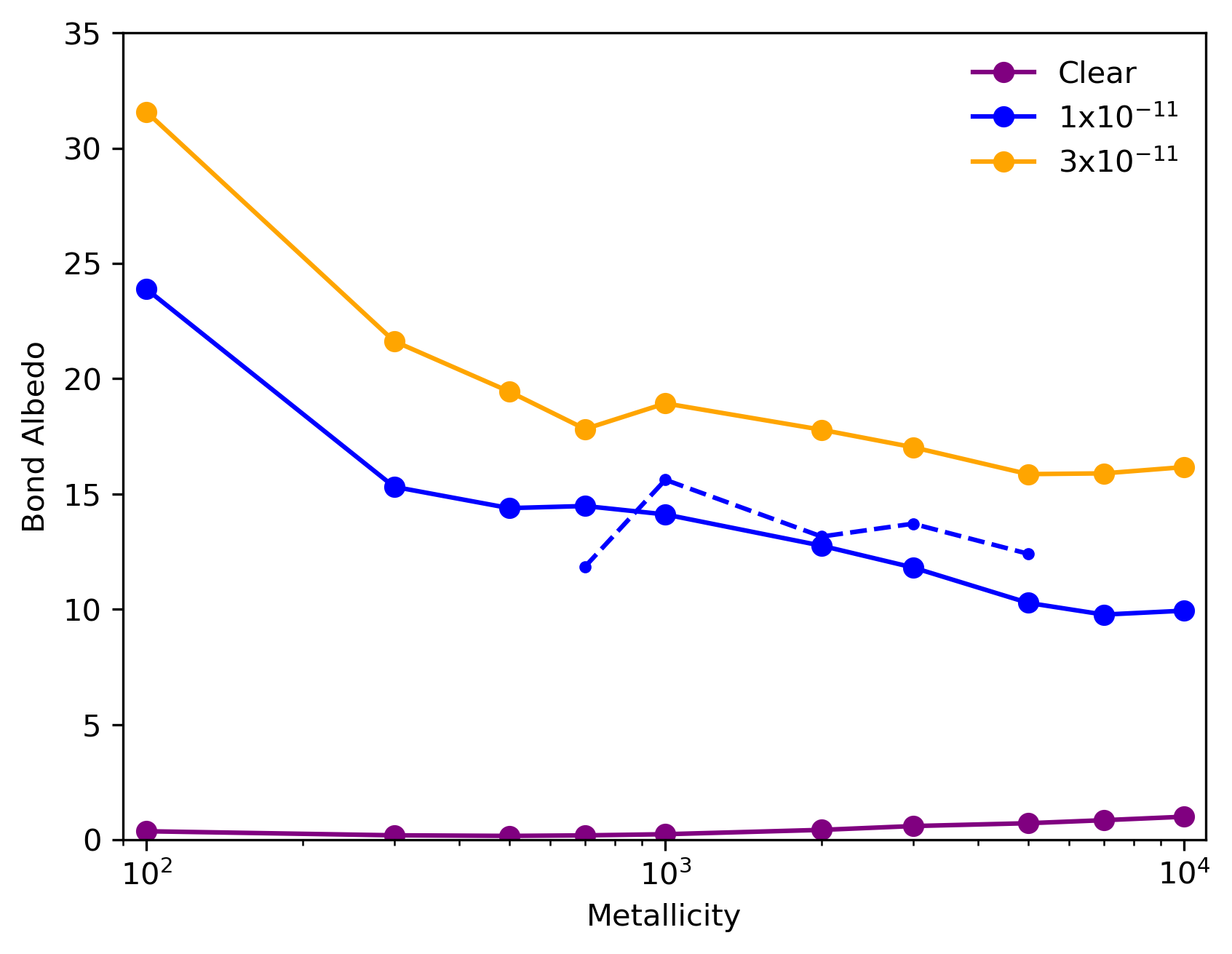}
\caption{Reduced $\chi^2$ of the JWST eclipse observations across metallicity (left) and the corresponding atmospheric bond albedo (right) for different scenarios of haze mass flux.  Dashed lines present the results for the high atmospheric mixing case (see sensitivity tests).}\label{chi2_eclipse}
\end{figure}

Our thermal structure simulations describe both the thermal and stellar radiation propagation in the atmosphere, thus directly provide spectra that can be compared with the JWST observations and inform on the Bond albedo. Inspection of the thermal emission spectra (Fig.~\ref{eclipse}) suggest that a clear atmosphere is not an optimal fit of the observations. This conclusion is further verified by the minimum reduced $\chi^2$ achieved, for 2000x solar metallicity, which is 2.4 (Fig.~\ref{chi2_eclipse}). Nevertheless, these simulations do demonstrate that metallicities between 1000x and 2000x solar are most consistent with the observed thermal emission, i.e. that the precision of the JWST eclipse observations can set improved constraints on the atmospheric metallicity of GJ1214 b.\rm

Including the photochemical hazes and clouds in the atmospheric structure simulations improves significantly the simulated thermal emission spectra relative to the observations (Fig.~\ref{eclipse}). The hazy scenarios provide improved reduced $\chi^2$ values relative to the clear atmosphere cases, with minima at $\chi^2$ $\sim$1.3. But while, the minimum $\chi^2$ achieved for the two haze scenarios is similar, the metallicity region implied for each haze mass flux is slightly different: we find reduced $\chi^2$ minima within 1000-2000x solar for the 1$\times$10$^{-11}$ g cm$^{-2}$s$^{-1}$ case and within 2000-3000x solar for the 3$\times$10$^{-11}$ g cm$^{-2}$s$^{-1}$ case. Nevertheless, what is important from this analysis is that the high metallicity end of the studied range results in worse fits of the observations, i.e. the eclipse observations continue to provide constraints for the atmospheric metallicity for the hazy scenarios as well. 

Previous estimates of GJ1214 b's Bond albedo suggest values of $\sim$40 - 50$\%$, assuming a single temperature black body behaviour to fit the JWST spectrum \cite{Kempton23}. Although such a wavelength dependence is representative at the observed wavelengths, our simulated spectra demonstrate that at slightly shorter wavelengths the $\lambda$-dependence of the thermal emission diverges significantly from a black body, thus making the Bond albedo estimation through such an approach approximate. We thus advance over the previous estimate by calculating directly the atmospheric Bond albedo through the reflected stellar light component of our simulations. Our calculated Bond albedo for clear atmospheres is systematical below 1$\%$ for all metallicity cases considered (Fig.~\ref{chi2_eclipse}). Certainly, a reflecting haze component is necessary to get higher albedo values. For the evaluated haze scenarios we find that the JWST observations are consistent with Bond albedos of $\sim$10-20$\%$. These high albedo values are dominantly driven by the haze scattering, while the cloud contribution is small due to the high pressures at which the cloud density attains significant values.

\begin{figure}
\includegraphics[scale=0.5]{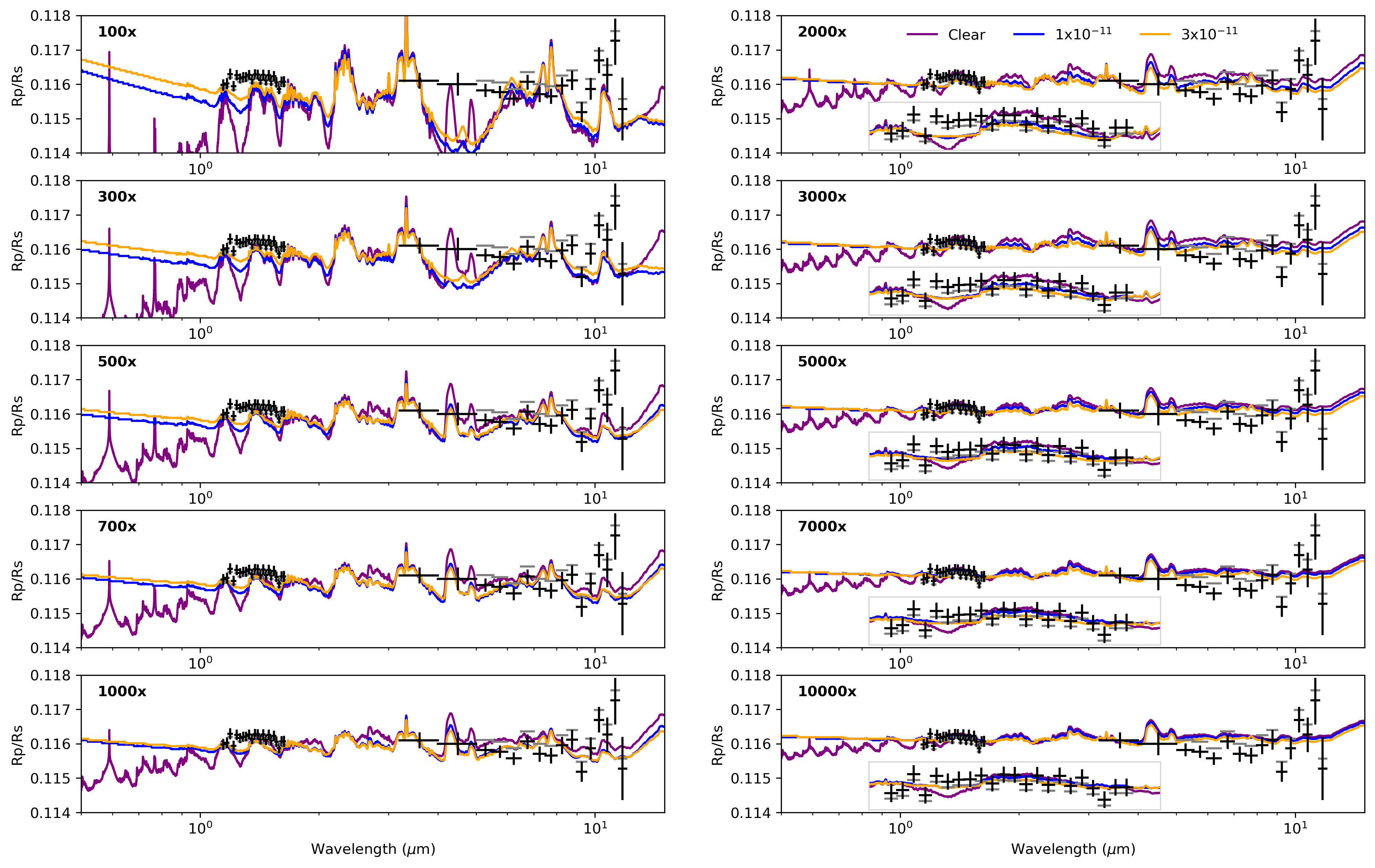}
\caption{Observed (points with error bars corresponding to one standard deviation) and simulated (lines) transit spectra of GJ 1214b for different metallicity scenarios (panels). Lines corresponds to clear atmosphere (purple), and the 10$^{-11}$ (blue) and 3$\times$10$^{-11}$ (orange) haze mass flux scenarios. Gray bars present the shift in the HST and JWST observations when their corresponding mean transit depth is scaled to the 3.6 $\mu$m Spitzer value. }\label{transit}
\end{figure}

We further consider the constraints from transit observations for the scenarios evaluated for the eclipse observations. For this evaluation we use observations from HST/WFC \cite{Kreidberg14}, from Spitzer \cite{Fraine13} and JWST/ \cite{Kempton23}. For the Spitzer data we use the results obtained with the \cite{Berta12} orbital parameters to have a consistent representation among the different data sets \cite{Gao23}. The simulated transit spectra demonstrate the need for both high metallicity and haze in the atmosphere of GJ1214 b (Fig.~\ref{transit}).  Increasing the metallicity makes the transit spectrum flatter but even at 10 000x solar metallicity haze is still required to fit the observations, particularly the observed flatness in the HST spectrum (see inset plots in Fig.~\ref{transit}). 

\begin{figure}
\includegraphics[scale=0.5]{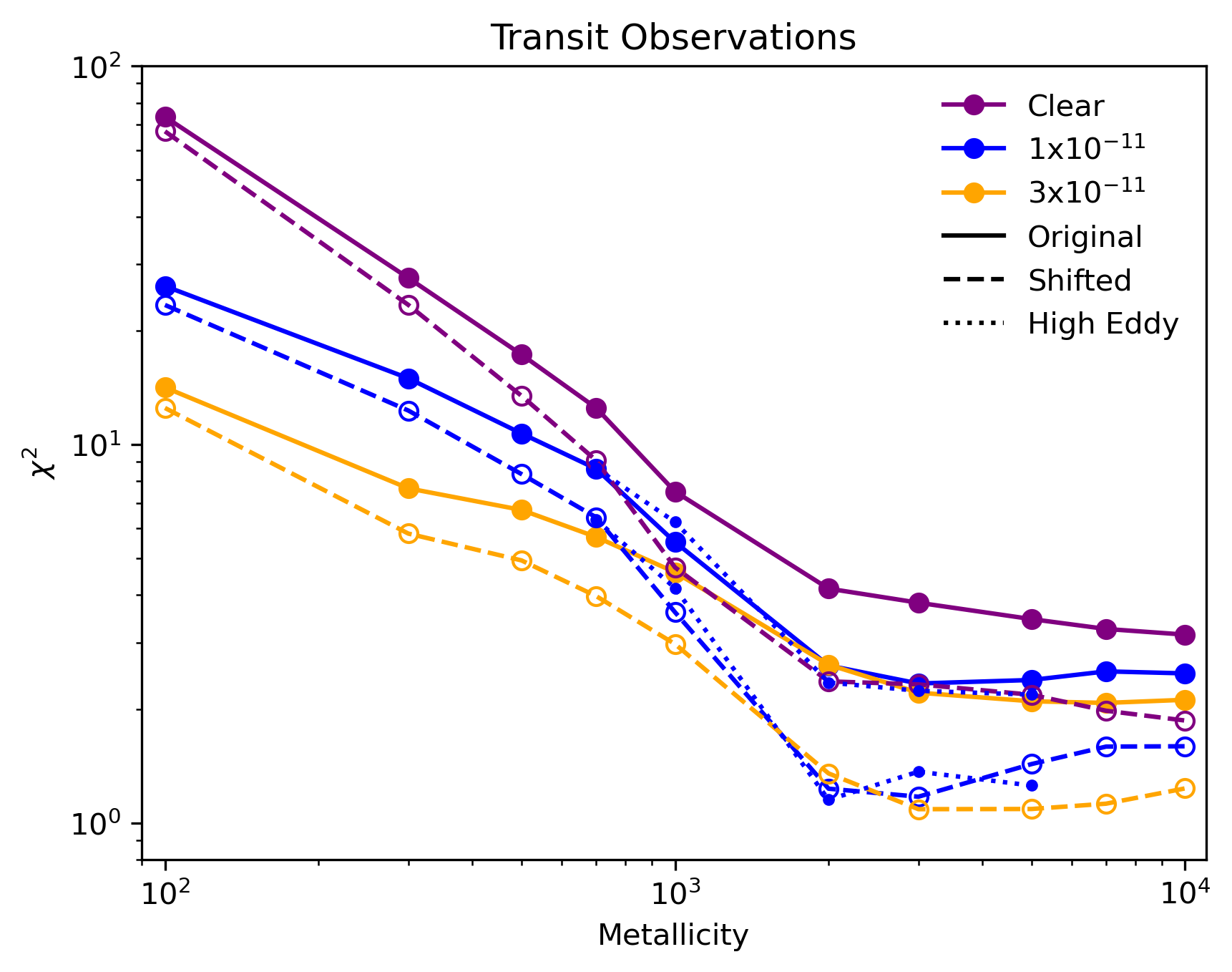}
\caption{Reduced $\chi^2$ for the transit observations of GJ1214b for different atmospheric scenarios. Solid lines correspond to the unshifted data and dashed to the re-normalized data to the 3.6 $\mu$m Spitzer observation. Dotted lines for the 10$^{-11}$ g cm$^{-2}$s$^{-1}$ haze mass flux case present the case of stronger atmospheric mixing (see sensitivity tests).}\label{chi2}
\end{figure}

The reduced $\chi^2$ for the two haze mass flux  scenarios decrease drastically with increasing metallicity and appear to approach a plateau above 2000 x solar, with the higher haze mass flux scenario providing a slightly better fit (Fig.~\ref{chi2}). At 10 000x solar metallicity the clear atmosphere scenario provides a reduced $\chi^2$ value of 3.1, significantly larger than the corresponding values of the two haze scenarios (2.5 for 1$\times$10$^{-11}$ g cm$^{-2}$s$^{-1}$ and 2.1 for 3$\times$10$^{-11}$ g cm$^{-2}$s$^{-1}$). Thus, the clear atmosphere scenario is eliminated from these results as well. Note that for the $\chi^2$ evaluation we considered all available observations at wavelengths, $\lambda<10 \mu$m, i.e. excluding only wavelengths from the JWST/MIRI LRS analysis for which the background is not well characterised \cite{Kempton23}. 

We also evaluated the corresponding $\chi^2$ for a scenario where the HST and JWST observations are adjusted to match the Spitzer 3.6 $\mu$m observation. This shift results in slightly smaller/larger transit depths for HST/JWST observations and its purpose is to evaluate the possible implications of temporal changes in the star output among the different observing times and how they affect the relative positioning of the three independent observations \cite{Gao23}. In this case we obtain smaller reduced $\chi^2$ values with minima near 1.2 for metallicities between 2000x and 5000x solar for the 1$\times$10$^{-11}$ g cm$^{-2}$s$^{-1}$ haze mass flux case and near 1.1 for metallicities between 3000x and 5000x solar for the 3$\times$10$^{-11}$ g cm$^{-2}$s$^{-1}$ haze mass flux case (see dashed lines in Fig.~\ref{chi2}). This improved match to the observations indicates that our simulated transit spectra are more consistent with a flat spectrum, relative to the unscaled observations. Vice versa, with the applied scaling we conclude that haze mass fluxes in the range considered (and with the specific optical properties assumed) are sufficient to explain the observations. Irrespective of the observation scaling however, we obtain qualitatively the same conclusions regarding the need for haze to explain the observations, as well as, the tendency for metallicities above 1000x solar. 

\begin{figure}
\includegraphics[scale=0.5]{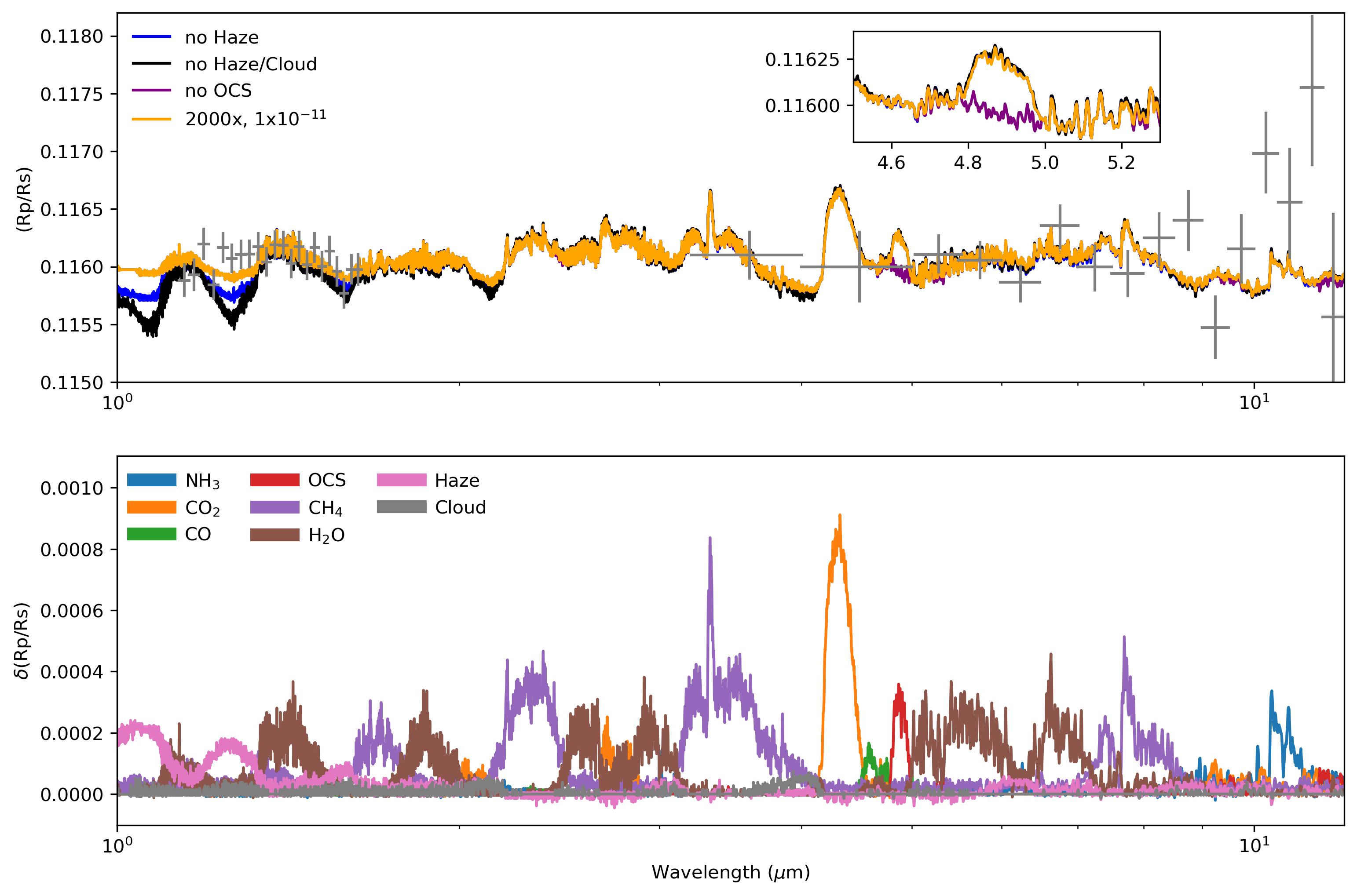}
\caption{Top: Sensitivity of transit spectrum for the 2000x solar metallicity case with 10$^{-11}$ haze mass flux to different opacity contributions. Bottom: Differential transit spectrum presenting the opacity contributions from different gases and haze.}\label{gases}
\end{figure}

Although the presence of haze tends to flatten the spectrum, its effect is limited to short wavelengths below 2$\mu$m (Fig.~\ref{gases}). Cloud particles have a small contribution in the transit spectrum relative to the haze. Without the haze, clouds would make the spectrum more shallow relative to the clear atmosphere spectrum, but their contribution would not be able to fill in the troughs in the water bands below 2 $\mu$m to the degree suggested by the observations. At longer wavelengths the gaseous opacities contribute to the flatness of the spectrum. Opacities from H$_2$O, CH$_4$, CO$_2$, CO, NH$_3$ have similar magnitudes and their combined effect cover the whole spectrum between 1 and 10 $\mu$m (Fig.~\ref{gases}). Thus, the cumulative opacity in tandem with the decreasing atmospheric scale height with increasing metallicity make the transit spectra rather flat at these wavelengths. Clouds also have a minor contribution in between the gas bands near 1.6, 2.1 and 4.1 $\mu$m, however their effect is small ($\delta$(Rp/Rs)$\sim$5$\times$10$^{-5}$). 

\begin{figure}
\includegraphics[scale=0.5]{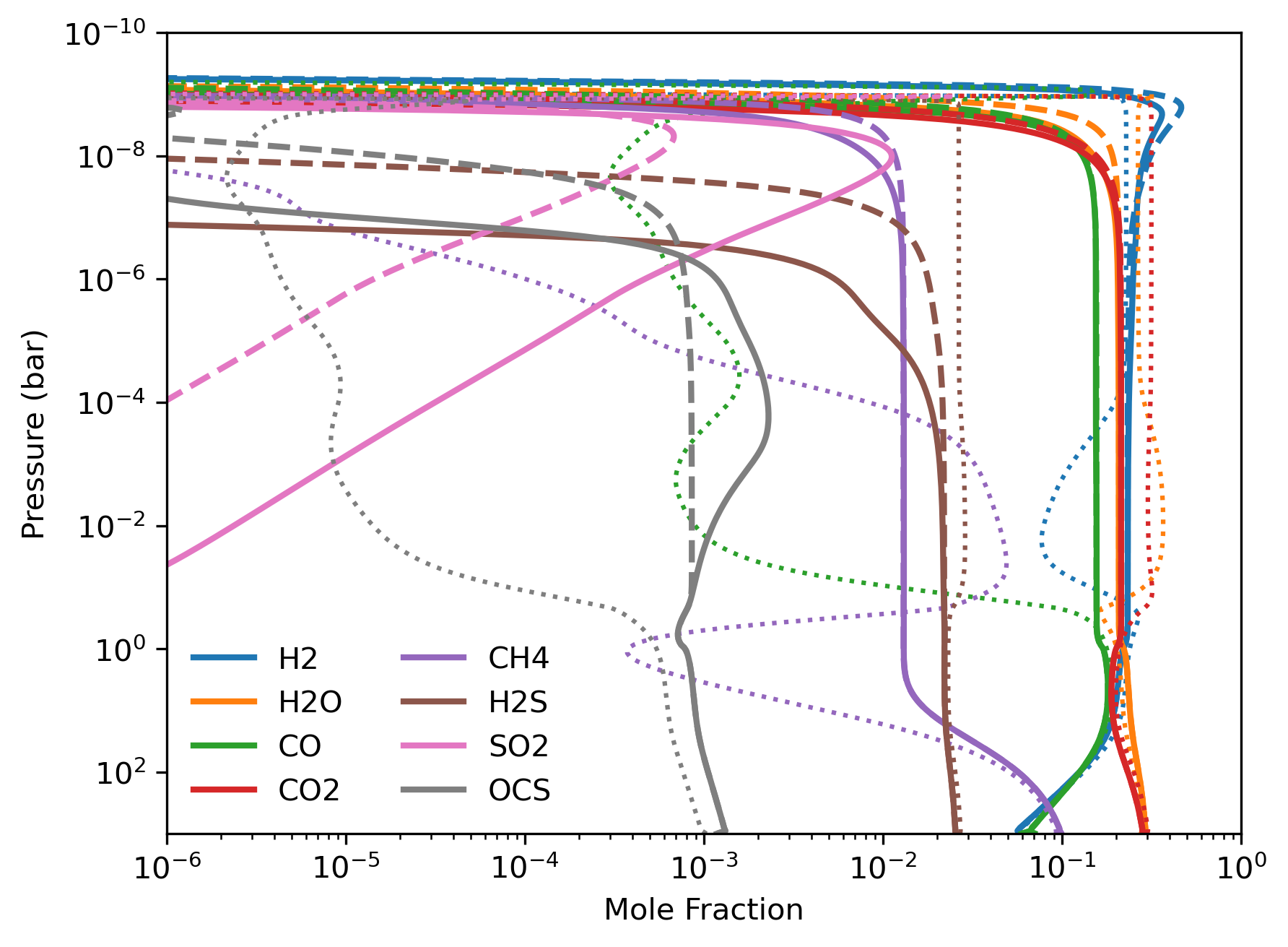}
\caption{Atmospheric composition for the 2000x solar metallicity case and the 10$^{-11}$ g cm$^{-2}$s$^{-1}$ haze mass flux case. Solid and dashed lines present the composition results for the disk and limb geometries, respectively. Dotted lines are the equilibrium composition results for the same temperature profile.}\label{COS}
\end{figure}

The gaseous abundances responsible for the transit flatness are significantly affected by disequilibrium, in two major aspects; first by the impact of the haze (a disequilibrium product itself) on the thermal structure, and second by the atmospheric mixing and photochemistry on the gas abundances (Fig.~\ref{COS}), with the latter having its own effect on the thermal structure. The self-consistent description used in this study is therefore indispensable for exploring these interactions. The abundances of the main opacity contributors, are significantly different compared to the equilibrium profiles at the same temperature, while photochemically produced SO$_2$ is orders of magnitudes higher than its equilibrium density (as detected also in hot-Jupiter WASP-39b \cite{Tsai23}), but for GJ1214 b its abundances is not significant at the transit probed pressures ($\sim$1mbar - 1$\mu$bar). On the contrary, OCS an abundant photochemical product of the organo-sulfur chemistry according to our calculations, has a clear signature between 4.8 and 5.0 $\mu$m in the simulated spectra and could potentially be observed in future observations (Fig.~\ref{gases}).

\begin{figure}
\centering
\includegraphics[scale=0.5]{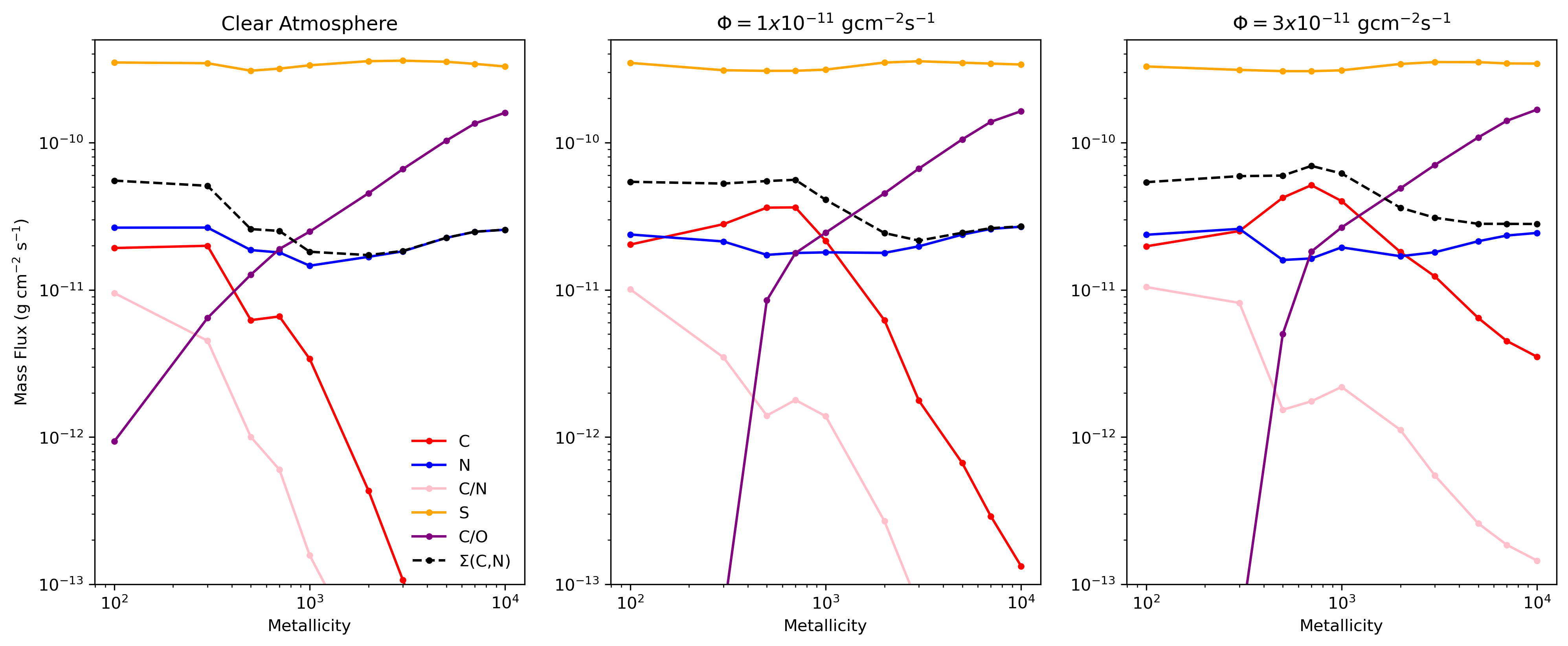}
\caption{Mass fluxes generated from the photolysis of different composition haze precursors, for the three haze mass flux cases considered.}\label{chemistry}
\end{figure}

The disequilibrium chemistry component of our simulations allows also to estimate the mass flux generated from the photolysis of major haze precursors in the atmosphere, which permits an evaluation for the capacity of the atmosphere to generate the required mass flux to explain the observations. We consider precursors of pure carbon (CH$_4$, C$_2$H$_2$, C$_2$H$_4$) and pure nitrogen (N$_2$, NH$_3$) composition, as well as, combined C/N (HCN, CH$_2$NH) precursors (Fig.~\ref{chemistry}). In addition, we evaluated the possible contribution from sulphur precursors (H$_2$S, S$_2$) and from C/O (CO, CO$_2$) compounds.  
The relative contributions of the various precursors change according to metallicity but also according to the assumed haze mass flux. While the former effect is anticipated and previously identified \cite{Lavvas19}, the latter effect occurs due to the direct impact of the haze opacity on the radiation field, as well as, the indirect modifications imposed on the gaseous abundances by the changes on the atmospheric thermal structure due to the presence of the haze. Our results reveal that haze formation modifies the atmospheric composition in such a way as to increase its capacity to produce haze, at least for the conditions tested. 
Overall, we find that mass fluxes of $\sim$(2-6)$\times$10$^{-11}$ g cm$^{-2}$ s$^{-1}$ are generated from all C and N composition precursors in the atmosphere, under all metallicity cases, suggesting that the atmosphere can generate the haze mass flux implied by the observations, with a rather large haze formation yield, though. On the other hand, sulfur precursors provide an additional contribution up to $\sim$3$\times$10$^{-10}$ g cm$^{-2}$ s$^{-1}$, indicating that photochemical hazes in sub-Neptune atmospheres could have a significant organosulfuric component. Laboratory experiments on the formation of photochemical hazes demonstrate that the presence of sulfur enhances the haze production at conditions similar to those identified here in the atmosphere of GJ1214b \cite{He20,Vuitton21}. However, we lack the optical properties of such type hazes to assist in their possible identification from observations. Studies on the irradiation of organic ices reveal that the inclusion of H$_2$S in the experimental process results in organosulfuric residues with multiple new absorption bands relative to the pure organic residues \cite{Mahjoub21, Mahjoub23}. Thus, it is possible that such features could be observable in future high-resolution transit observations of this or similar exoplanets.

\section{Appendix}

\subsection{Model description}
We evaluate the atmospheric thermal structure assuming the GJ1214  stellar flux with complete energy redistribution \cite{Lavvas19,Lavvas21}, and a low (T$_{int}$ = 30K) intrinsic temperature \cite{Gao23}. The lower boundary (at 1000 bar) conditions of atmospheric composition are based on a thermochemical equilibrium model \cite{Woitke18}. Atmospheric mixing is taken into account through an eddy profile (K$_{ZZ}$). Atmospheric metallicity defines the mean molecular weight of the atmosphere, which has implications for the mixing efficiency as demonstrated by GCM simulations \cite{Charnay15a}. However, eddy profile estimates from GCMs are subject to uncertainties of factors between 10 and 100 \cite{Parmentier13}, while hazes can have a strong impact on the atmospheric thermal structure and dynamics, therefore on the retrieved mixing efficiency \cite{Steinrueck23}. We thus consider a common eddy profile for all studied cases, to minimise our free parameters, but we evaluate how the mixing magnitude could impact our results.

We evaluate the size distribution of haze particles following their growth through coagulation and their transport by settling and atmospheric mixing \cite{Lavvas17}. Production is assumed to occur through a gaussian profile centred at 1$\mu$bar and with an initial radius of 1 nm. We evaluate the haze optical properties based on refractive indices obtained from exoplanet haze analogs \cite{He23} produced from a H/C/N/O gas mixture at temperature conditions similar to those anticipated for the atmosphere of GJ1214 b. We calculate the cloud size distributions considering KCL, NaCL and ZnS condensates that are expected to form at the simulated temperature conditions. Although Na is drastically reduced due to the formation of Na$_2$S in the deeper atmosphere, thermochemical equilibrium simulations \cite{Woitke18} suggest that there is a residual Na remaining in the form of gaseous NaCL that can thus condense if the temperature conditions permit it. KCL and NaCL condensation proceeds as a phase transition from the gas phase components to the condensates, while ZnS forms a non-molecular solid, therefore we treat Zn (that has a lower abundance than S) as an indicator of its condensation with a corresponding saturation vapour pressure from \cite{Morley15}. Condensates nucleate at the surface of photochemical haze particles (heterogeneous nucleation). This approach has recently been applied in Wasp-39b's atmosphere, where the formation of Na$_2$S clouds on the surface of photochemical haze particles allowed for the interpretation of HST and JWST observations from UV to IR \cite{Arfaux24}. The interaction of the anticipated condensates on the surface of photochemical hazes is unknown, thus we treat this interaction as a free parameter by assuming an efficient nucleation environment (small contract angle) and evaluating how different haze mass fluxes would affect the cloud properties. We note that the goal of this exercise is to evaluate the optimal conditions under which clouds could interpret the observations, which explains the efficient nucleation assumption considered.

The simulated thermal structure, dis-equilibrium composition and haze/cloud size distributions for the disk average conditions are then used for simulating the eclipse observations. For the transit observations we further evaluate the dis-equilibrium composition for limb geometry assuming the disk-average thermal structure. Certainly atmospheres are variable in 3D. The advantage of the detailed 1D study presented here is that it can quickly limit the phase space of conditions that should be explored by the more time-demanding 3D model, as well as, provide results for the metallicity and the degree of cloud/haze contribution in the observations. We additionally explore below how the assumed temperature profiles at the terminators could affect our conclusions. 

\subsection{Sensitivity tests}
\begin{figure}
\includegraphics[scale=0.5]{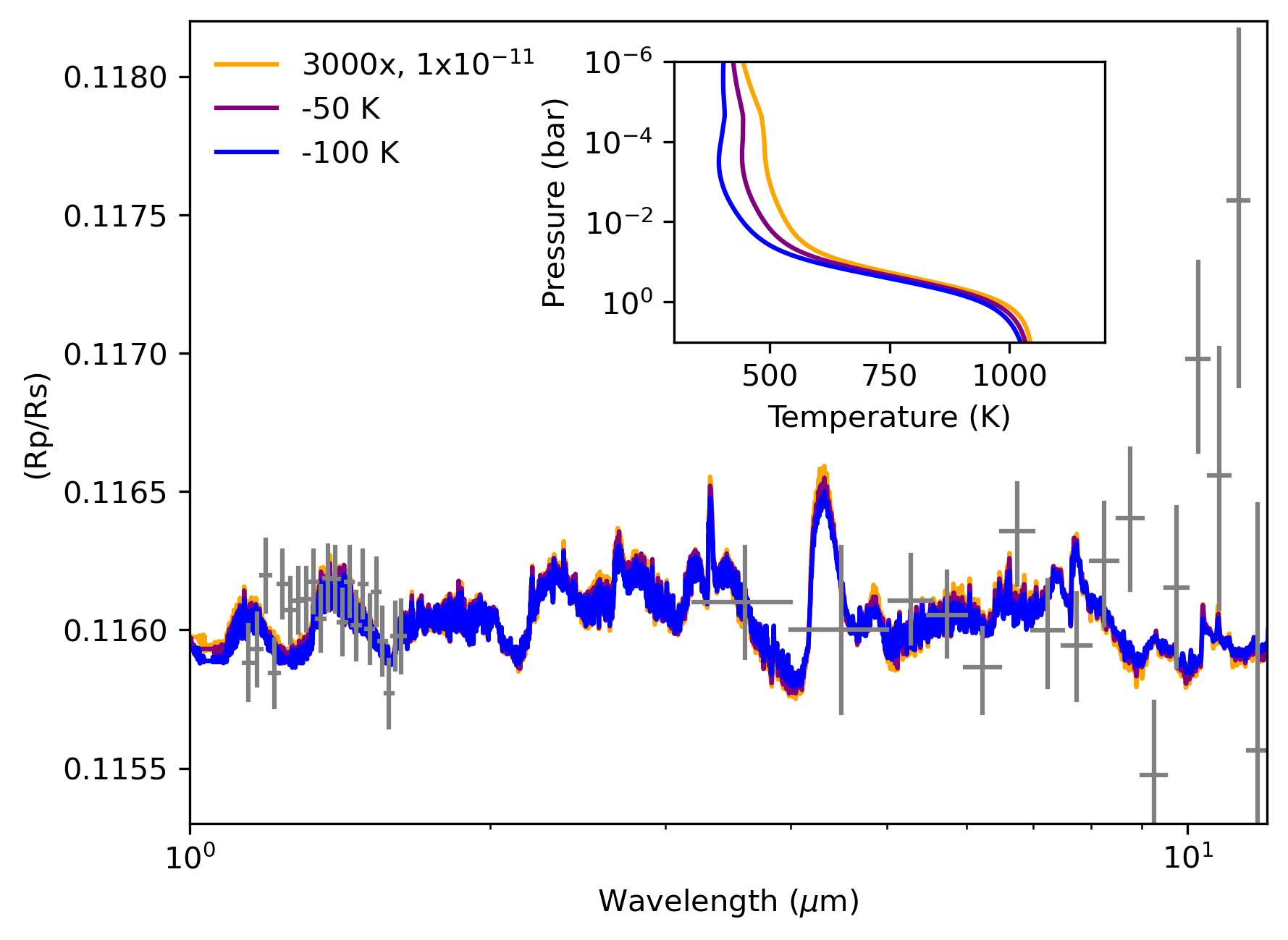}
\caption{Sensitivity of transit spectrum on temperature variations of the terminator.}\label{sense}
\end{figure}

Transit observations probe the planetary terminators that probably are at different temperature conditions than those of the disk assumed so far. To evaluate the impact of our assumption on the transit spectra we evaluated the atmospheric composition and haze $\&$ cloud particle properties assuming variations of the thermal structure where the profile is cooled by a gaussian shape temperature decrease centred at 1 mbar and with magnitudes of 50 or 100 K. Such a shape for the temperature modification is motivated by GCM simulations of this planet (M. Steinrueck personal communication). Our results however, indicate that the impact of the temperature cooling on the dis-equilibrium chemical composition of the gases and the haze distribution is too small to modify our previous conclusions (Fig.~\ref{sense}). On the other hand re-distribution of the particles due to the atmospheric circulation may have a more important impact that needs to be further investigated with GCM simulations \cite{Steinrueck23}.

We also considered a scenario of a 10$\times$ stronger atmospheric mixing compared to the nominal case used previously, to evaluated how eddy mixing variations may affect our results. We explore this case for metallicities between 700 and 5000$\times$ solar that are most consistent with the observations and for the 1$\times$10$^{-11}$ g cm$^{-2}$s$^{-1}$ haze mass flux case. We find that despite the changes in the thermal structure, composition, and particle size distributions imposed by the enhanced mixing, the metallicity constraints from both eclipse and transit observations, as well as the Bond albedo evaluation remain unaffected (see dashed lines in Figs.~\ref{chi2_eclipse} $\&$ \ref{chi2}). We thus conclude that the constraints we obtain on the atmospheric characterisation of GJ1214b are independent of these variations on temperature and mixing efficiency.

\section{Data availability}
Data sets generated during the current study are available from the corresponding author on reasonable request. 

\section{Code availability}
The methods used on this study where previously reviewed and published \cite{Arfaux22,Arfaux24}. The code is not publicly released as it is not yet in a user-friendly design to facilitate broad use.

\section{Author contributions}
PL designed and executed the research. SP explored aspects of cloud formation, identified required input and run simulations. AA explored equilibrium composition for the boundary conditions and run simulations. PL wrote the manuscript and all authors provided comments. 

\section{Acknowledgments}
Part of the work was performed during the internship of SP at GSMA/France under the support of the Erasmus+ program. The work was further supported by the Programme National de Planetologie of INSU/CNRS under the project TISSAGE.

\bibliographystyle{ieeetr} 
\bibliography{refs}

\begin{thebibliography}{10}

\bibitem{Owen17}
J.~E. {Owen} and Y.~{Wu}, ``{The Evaporation Valley in the Kepler Planets},''
  {\em The Astrophysical Journal}, vol.~847, p.~29, Sept. 2017.

\bibitem{Kite20}
E.~S. Kite and M.~N. Barnett, ``Exoplanet secondary atmosphere loss and
  revival,'' {\em Proceedings of the National Academy of Sciences}, vol.~117,
  no.~31, pp.~18264--18271, 2020.

\bibitem{Koskinen22}
T.~T. {Koskinen}, P.~{Lavvas}, C.~{Huang}, G.~{Bergsten}, R.~B. {Fernandes},
  and M.~E. {Young}, ``{Mass Loss by Atmospheric Escape from Extremely Close-in
  Planets},'' {\em The Astrophysical Journal}, vol.~929, p.~52, Apr. 2022.

\bibitem{Kreidberg14}
L.~Kreidberg, J.~L. Bean, J.-M. D{\'e}sert, B.~Benneke, D.~Deming, K.~B.
  Stevenson, S.~Seager, Z.~Berta-Thompson, A.~Seifahrt, and D.~Homeier,
  ``{Clouds in the atmosphere of the super-Earth exoplanet GJ1214b},'' {\em
  Nature}, vol.~505, pp.~69--72, Jan. 2014.

\bibitem{Narita13}
N.~{Narita}, A.~{Fukui}, M.~{Ikoma}, Y.~{Hori}, K.~{Kurosaki}, Y.~{Kawashima},
  T.~{Nagayama}, M.~{Onitsuka}, A.~{Sukom}, Y.~{Nakajima}, M.~{Tamura},
  D.~{Kuroda}, K.~{Yanagisawa}, T.~{Hirano}, K.~{Kawauchi}, M.~{Kuzuhara},
  H.~{Ohnuki}, T.~{Suenaga}, Y.~H. {Takahashi}, H.~{Izumiura}, N.~{Kawai}, and
  M.~{Yoshida}, ``{Multi-color Transit Photometry of GJ 1214b through BJHK
  $_{s}$ Bands and a Long-term Monitoring of the Stellar Variability of GJ
  1214},'' {\em The Astrophysical Journal}, vol.~773, p.~144, Aug. 2013.

\bibitem{Nascimbeni15}
V.~Nascimbeni, M.~Mallonn, G.~Scandariato, I.~Pagano, G.~Piotto, G.~Micela,
  S.~Messina, G.~Leto, K.~G. Strassmeier, S.~Bisogni, and R.~Speziali, ``{Large
  Binocular Telescope view of the atmosphere of GJ1214b},'' {\em Astronomy and
  Astrophysics}, vol.~579, p.~A113, July 2015.

\bibitem{Fraine13}
J.~D. Fraine, D.~Deming, M.~Gillon, E.~Jehin, B.-O. Demory, B.~Benneke,
  S.~Seager, N.~K. Lewis, H.~Knutson, and J.-M. D{\'e}sert, ``{Spitzer Transits
  of the Super-Earth GJ1214b and Implications for its Atmosphere},'' {\em The
  Astrophysical Journal}, vol.~765, p.~127, Mar. 2013.

\bibitem{Morley13}
C.~V. Morley, J.~J. Fortney, E.~M.~R. Kempton, M.~S. Marley, C.~Visscher, and
  K.~Zahnle, ``{Quantitatively Assessing the Role of Clouds in the Transmission
  Spectrum of GJ 1214b},'' {\em The Astrophysical Journal}, vol.~775, p.~33,
  Sept. 2013.

\bibitem{Morley15}
C.~V. Morley, J.~J. Fortney, M.~S. Marley, K.~Zahnle, M.~Line, E.~Kempton,
  N.~Lewis, and K.~Cahoy, ``{Thermal Emission and Reflected Light Spectra of
  Super Earths with Flat Transmission Spectra},'' {\em The Astrophysical
  Journal}, vol.~815, p.~110, Dec. 2015.

\bibitem{Charnay15a}
B.~Charnay, V.~Meadows, and J.~Leconte, ``{3D Modeling of GJ1214b's Atmosphere:
  Vertical Mixing Driven by an Anti-Hadley Circulation},'' {\em The
  Astrophysical Journal}, vol.~813, p.~15, Nov. 2015.

\bibitem{Gao18}
P.~Gao and B.~Benneke, ``{Microphysics of KCl and ZnS Clouds on GJ1214b},''
  {\em The Astrophysical Journal}, vol.~863, p.~23, Aug. 2018.

\bibitem{Ohno18}
K.~Ohno and S.~Okuzumi, ``{Microphysical Modeling of Mineral Clouds in GJ1214 b
  and GJ436 b: Predicting Upper Limits on the Cloud-top Height},'' {\em The
  Astrophysical Journal}, vol.~859, p.~34, May 2018.

\bibitem{Kawashima18}
Y.~Kawashima and M.~Ikoma, ``{Theoretical Transmission Spectra of Exoplanet
  Atmospheres with Hydrocarbon Haze: Effect of Creation, Growth, and Settling
  of Haze Particles. I. Model Description and First Results},'' {\em The
  Astrophysical Journal}, vol.~853, p.~7, Jan. 2018.

\bibitem{Adams19}
D.~Adams, P.~Gao, I.~de~Pater, and C.~V. Morley, ``{Aggregate Hazes in
  Exoplanet Atmospheres},'' {\em The Astrophysical Journal}, vol.~874:61,
  p.~14pp, Mar. 2019.

\bibitem{Lavvas19}
P.~Lavvas, T.~Koskinen, M.~E. Steinrueck, A.~G. Mu{\~n}oz, and A.~P. Showman,
  ``{Photochemical Hazes in Sub-Neptunian Atmospheres with a Focus on GJ
  1214b},'' {\em The Astrophysical Journal}, vol.~878, p.~16, June 2019.

\bibitem{Kempton23}
E.~M.~R. Kempton, M.~Zhang, J.~L. Bean, M.~E. Steinrueck, A.~A.~A. Piette,
  V.~Parmentier, I.~Malsky, M.~T. Roman, E.~Rauscher, P.~Gao, T.~J. Bell,
  Q.~Xue, J.~Taylor, A.~B. Savel, K.~E. Arnold, M.~C. Nixon, K.~B. Stevenson,
  M.~Mansfield, S.~Kendrew, S.~Zieba, E.~Ducrot, A.~Dyrek, P.-O. Lagage, K.~G.
  Stassun, G.~W. Henry, T.~Barman, R.~Lupu, M.~Malik, T.~Kataria, J.~Ih, G.~Fu,
  L.~Welbanks, and P.~McGill, ``{A reflective, metal-rich atmosphere for GJ
  1214b from its JWST phase curve},'' {\em Nature}, vol.~620, no.~7972,
  pp.~67--71, 2023.

\bibitem{Gao23}
P.~Gao, A.~A.~A. Piette, M.~E. Steinrueck, M.~C. Nixon, M.~Zhang, E.~M.~R.
  Kempton, J.~L. Bean, E.~Rauscher, V.~Parmentier, N.~E. Batalha, A.~B. Savel,
  K.~E. Arnold, M.~T. Roman, I.~Malsky, and J.~Taylor, ``{The Hazy and
  Metal-Rich Atmosphere of GJ 1214 b Constrained by Near and Mid-Infrared
  Transmission Spectroscopy},'' {\em The Astrophysical Journal}, vol.~951,
  no.~2, p.~16, 2023.

\bibitem{Arfaux22}
A.~Arfaux and P.~Lavvas, ``{A large range of haziness conditions in hot-Jupiter
  atmospheres},'' {\em MNRAS}, vol.~515, pp.~4753--4779, July 2022.

\bibitem{Lodders19b}
K.~Lodders, ``The chemical composition of the solar system,'' in {\em Nuclei in
  the Cosmos XV} (A.~Formicola, M.~Junker, L.~Gialanella, and G.~Imbriani,
  eds.), (Cham), pp.~165--170, Springer International Publishing, 2019.

\bibitem{Berta12}
Z.~K. {Berta}, D.~{Charbonneau}, J.-M. {D{\'e}sert}, E.~{Miller-Ricci Kempton},
  P.~R. {McCullough}, C.~J. {Burke}, J.~J. {Fortney}, J.~{Irwin}, P.~{Nutzman},
  and D.~{Homeier}, ``{The Flat Transmission Spectrum of the Super-Earth
  GJ1214b from Wide Field Camera 3 on the Hubble Space Telescope},'' {\em The
  Astrophysical Journal}, vol.~747, p.~35, Mar. 2012.

\bibitem{Tsai23}
S.-M. {Tsai}, E.~K.~H. {Lee}, D.~{Powell}, P.~{Gao}, X.~{Zhang}, J.~{Moses},
  E.~{H{\'e}brard}, O.~{Venot}, V.~{Parmentier}, S.~{Jordan}, R.~{Hu}, M.~K.
  {Alam}, L.~{Alderson}, N.~M. {Batalha}, J.~L. {Bean}, B.~{Benneke}, C.~J.
  {Bierson}, R.~P. {Brady}, L.~{Carone}, A.~L. {Carter}, K.~L. {Chubb},
  J.~{Inglis}, J.~{Leconte}, M.~{Line}, M.~{L{\'o}pez-Morales}, Y.~{Miguel},
  K.~{Molaverdikhani}, Z.~{Rustamkulov}, D.~K. {Sing}, K.~B. {Stevenson}, H.~R.
  {Wakeford}, J.~{Yang}, K.~{Aggarwal}, R.~{Baeyens}, S.~{Barat}, M.~{de
  Val-Borro}, T.~{Daylan}, J.~J. {Fortney}, K.~{France}, J.~M. {Goyal},
  D.~{Grant}, J.~{Kirk}, L.~{Kreidberg}, A.~{Louca}, S.~E. {Moran},
  S.~{Mukherjee}, E.~{Nasedkin}, K.~{Ohno}, B.~V. {Rackham}, S.~{Redfield},
  J.~{Taylor}, P.~{Tremblin}, C.~{Visscher}, N.~L. {Wallack}, L.~{Welbanks},
  A.~{Youngblood}, E.-M. {Ahrer}, N.~E. {Batalha}, P.~{Behr}, Z.~K.
  {Berta-Thompson}, J.~{Blecic}, S.~L. {Casewell}, I.~J.~M. {Crossfield},
  N.~{Crouzet}, P.~E. {Cubillos}, L.~{Decin}, J.-M. {D{\'e}sert}, A.~D.
  {Feinstein}, N.~P. {Gibson}, J.~{Harrington}, K.~{Heng}, T.~{Henning},
  E.~M.~R. {Kempton}, J.~{Krick}, P.-O. {Lagage}, M.~{Lendl}, J.~D.
  {Lothringer}, M.~{Mansfield}, N.~J. {Mayne}, T.~{Mikal-Evans}, E.~{Palle},
  E.~{Schlawin}, O.~{Shorttle}, P.~J. {Wheatley}, and S.~N. {Yurchenko},
  ``{Photochemically produced SO$_{2}$ in the atmosphere of WASP-39b},'' {\em
  Nature Astronomy}, vol.~617, pp.~483--487, May 2023.

\bibitem{He20}
C.~He, S.~M. H{\"o}rst, N.~K. Lewis, X.~Yu, J.~I. Moses, P.~McGuiggan, M.~S.
  Marley, E.~M.~R. Kempton, S.~E. Moran, C.~V. Morley, and V.~Vuitton,
  ``Sulfur-driven haze formation in warm co2-rich exoplanet atmospheres,'' {\em
  Nature Astronomy}, vol.~4, no.~10, pp.~986--993, 2020.

\bibitem{Vuitton21}
V.~{Vuitton}, S.~E. {Moran}, C.~{He}, C.~{Wolters}, L.~{Flandinet}, F.-R.
  {Orthous-Daunay}, J.~I. {Moses}, J.~A. {Valenti}, N.~K. {Lewis}, and S.~M.
  {H{\"o}rst}, ``{H$_{2}$SO$_{4}$ and Organosulfur Compounds in Laboratory
  Analogue Aerosols of Warm High-metallicity Exoplanet Atmospheres},'' {\em The
  Planetary Science Journal}, vol.~2, p.~2, Feb. 2021.

\bibitem{Mahjoub21}
A.~{Mahjoub}, M.~E. {Brown}, M.~J. {Poston}, R.~{Hodyss}, B.~L. {Ehlmann},
  J.~{Blacksberg}, M.~{Choukroun}, J.~M. {Eiler}, and K.~P. {Hand}, ``{Effect
  of H$_{2}$S on the Near-infrared Spectrum of Irradiation Residue and
  Applications to the Kuiper Belt Object (486958) Arrokoth},'' {\em The
  Astrophysical Journal Letters}, vol.~914, p.~L31, June 2021.

\bibitem{Mahjoub23}
A.~Mahjoub, K.~Altwegg, M.~J. Poston, M.~Rubin, R.~Hodyss, M.~Choukroun, B.~L.
  Ehlmann, N.~H{\"a}nni, M.~E. Brown, J.~Blacksberg, J.~M. Eiler, and K.~P.
  Hand, ``Complex organosulfur molecules on comet 67p: Evidence from the rosina
  measurements and insights from laboratory simulations,'' {\em Science
  Advances}, vol.~9, no.~23, p.~eadh0394, 2023.

\bibitem{Lavvas21}
P.~Lavvas and A.~Arfaux, ``{Impact of photochemical hazes and gases on
  exoplanet atmospheric thermal structure},'' {\em MNRAS}, vol.~502, no.~4,
  pp.~5643--5657, 2021.

\bibitem{Woitke18}
P.~Woitke, C.~Helling, G.~H. Hunter, J.~D. Millard, G.~E. Turner, M.~Worters,
  J.~Blecic, and J.~W. Stock, ``{Equilibrium chemistry down to 100 K. Impact of
  silicates and phyllosilicates on the carbon to oxygen ratio},'' {\em
  Astronomy and Astrophysics}, vol.~614, p.~A1, June 2018.

\bibitem{Parmentier13}
V.~Parmentier, A.~P. Showman, and Y.~Lian, ``{3D mixing in hot Jupiters
  atmospheres. I. Application to the day/night cold trap in HD 209458b},'' {\em
  Astronomy and Astrophysics}, vol.~558, p.~A91, Oct. 2013.

\bibitem{Steinrueck23}
M.~E. {Steinrueck}, T.~{Koskinen}, P.~{Lavvas}, V.~{Parmentier}, S.~{Zieba},
  X.~{Tan}, X.~{Zhang}, and L.~{Kreidberg}, ``{Photochemical Hazes Dramatically
  Alter Temperature Structure and Atmospheric Circulation in 3D Simulations of
  Hot Jupiters},'' {\em The Astrophysical Journal}, vol.~951, p.~117, July
  2023.

\bibitem{Lavvas17}
P.~Lavvas and T.~Koskinen, ``{Aerosol Properties of the Atmospheres of
  Extrasolar Giant Planets},'' {\em The Astrophysical Journal}, vol.~847,
  p.~20, Sept. 2017.

\bibitem{He23}
C.~He, M.~Radke, S.~E. Moran, S.~M. Horst, N.~K. Lewis, J.~I. Moses, M.~S.
  Marley, N.~E. Batalha, E.~M.~R. Kempton, C.~V. Morley, J.~A. Valenti, and
  V.~Vuitton, ``{Optical Properties of Organic Hazes in Water-rich Exoplanet
  Atmospheres: Implications for Observations with JWST},'' {\em Nature
  Astronomy}, vol.~8, pp.~182--192, 2024.

\bibitem{Arfaux24}
A.~{Arfaux} and P.~{Lavvas}, ``{Coupling haze and cloud microphysics in
  WASP-39b's atmosphere based on JWST observations},'' {\em Monthly Notices of
  the Royal Astronomical Society}, vol.~530, pp.~482--500, May 2024.

\end{thebibliography}

\end{document}